\documentstyle[12pt,fleqn]{article}
\begin{document}
\begin{titlepage}
\vspace*{-62pt}
\begin{flushright}
UCLAHEP-92-44\\
DART-HEP-92/06\\
Revised Sept. 1993
\end{flushright}

\vspace{0.75in}
\centerline{\bf KINETICS OF SUB-CRITICAL BUBBLES AND THE ELECTROWEAK
TRANSITION}
\vskip 0.5cm
\centerline{Graciela Gelmini$^a$ and Marcelo Gleiser$^b$ }

\vskip 0.5 cm
\centerline{\it $^a$ Department of Physics,}
\centerline{\it University of California, Los Angeles, CA 90024}
\vskip 0.5cm
\centerline{$^b$\it Department of Physics and Astronomy,}
\centerline{\it Dartmouth College, Hanover, NH 03755}

\baselineskip=12pt

\vskip 1.5 cm
\centerline{\bf ABSTRACT}
\vskip 0.5 cm
\begin{quote}
We investigate the role of large amplitude
sub-critical thermal fluctuations
in the dynamics of first order phase transitions.
In particular, we obtain a kinetic equation for
the number density of sub-critical fluctuations of the broken-symmetric phase
within the symmetric phase,
modeled as spherical bubbles, and solve it analytically
for temperatures above the critical
temperature.
We study the approach to
equilibrium and obtain the equilibrium distribution of sub-critical bubbles
of the unstable phase
by examining three possible mechanisms responsible for their removal;
their shrinking, their coupling to thermal noise, and by thermal
fluctuations of the true vacuum inside them.
We show that for sufficiently strong transitions, either
the shrinking or the coupling to thermal noise dominate the dynamics.
As the strength of the transition weakens we show that sub-critical
fluctuations become progressively more important, as a larger
fraction of the total volume is occupied by the broken-symmetric phase,
until the point where
our analytical approach breaks down.  Our investigation
suggests that pre-transitional
phenomena may considerably change the dynamics of sufficiently weak
first-order transitions. We apply our results to
the standard electroweak transition.

\end{quote}

\vspace{0.1in}
e-mail: gelmini@physics.ucla.edu; gleiser@peterpan.dartmouth.edu

\end{titlepage}

\def\la{\mathrel{\mathpalette\fun <}}
\def\ga{\mathrel{\mathpalette\fun >}}
\def\fun#1#2{\lower3.6pt\vbox{\baselineskip0pt\lineskip.9pt
        \ialign{$\mathsurround=0pt#1\hfill##\hfil$\crcr#2\crcr\sim\crcr}}}
\def\mpl{{m_{Pl}}}
\def\f{\phi}
\def\s{\sigma}
\def\l{\lambda}
\def\rf{\langle \f \rangle}
\def\t{\theta}
\def\n{n(R,t)}
\def\ksection{\arabic{section}}
\def\theequation{\ksection.\arabic{equation}}
\def\thesection{}

\def\beq{\begin{equation}}
\def\eeq{\end{equation}}
\def\ba{\begin{eqnarray}}
\def\ea{\end{eqnarray}}
\def\re#1{{$^{\ref{#1}}$}}

\baselineskip=24pt
\thesection{\bf 1. INTRODUCTION}
\setcounter{section}{1}
\setcounter{equation}{0}
\vspace{24pt}

The study of the kinetics of first order phase transitions is by no means
a new topic. Since the pioneer work of Becker and D\"oring, the condensation
of supersatured vapor has been understood to occur by the thermal nucleation
of droplets of the liquid phase which, when larger than a critical size, will
grow and coalesce completing the transition.\re{NUC} Despite all the effort
dedicated to the study of the kinetics of phase transitions, their
intrinsic non-linear nature makes progress slow and restricted to simple
systems. Even for simple systems in the same universality class of the
Ising model many questions remain unanswered, from a microscopic understanding
of coarse-graining to the description of the late stages of the
transition.\re{LAN}


Our main interest is in the study of cosmological first order
transitions, where the cooling is enforced adiabatically by the expanding
Universe.\re{A}
In particular, we are only interested in
transitions where the expansion rate of the
Universe is much slower than typical fluctuation time-scales in the system,
{\it i.e.}, when $M_{Pl}\gg m$, where $M_{Pl}$ is the Planck mass, and $m$ is
the characteristic mass scale in the model.

In this work we will study analytically the early stages of a first
order phase transition.
That is, we will restrict our study to temperatures
above the critical
temperature $T_c$, which is defined as the temperature at which the homogeneous
part of the coarse-grained free energy density (the effective potential to some
order in perturbation theory) exhibits two degenerate minima. By focusing on
temperatures above the critical temperature, we will be examining the behavior
of thermal fluctuations around the high temperature equilibrium state of the
system. In particular, we will consider  models described by a potential
$V(\f,T)$ which at high temperatures has a minimum at $\rf=0$ and that, as the
temperature drops to some value $T_1$, develop a new minimum at
$\rf=\f_+(T)$.
At the critical temperature $T_c<T_1$, $V(0,T_c)=V(\f_+,T_c)$. We will examine
thermal fluctuations around $\rf=0$ as the temperature drops below $T_1$.

The reader may be wondering why should anyone be interested in studying the
kinetics of first order phase transitions above $T_c$. After all, it is a
well-known fact that first order transitions evolve through the nucleation
and subsequent percolation of
bubbles larger than a critical size as the temperature
drops below $T_c$.  This is indeed the
case for sufficiently strong transitions. However, for weak enough transitions
the critical bubble picture must be modified, as has been observed for several
condensed matter systems in the past two decades.  A particularly striking
example is the isotropic-nematic transition in certain
liquid crystals; the transition
is first order (there is a discontinuous jump of the order parameter)
even though there is no release of latent heat.\re{LQ} In fact, large amplitude
fluctuations of the nematic phase within the isotropic phase have been observed
{\it above} the critical temperature for the transition, a typical example of
the so-called pre-transitional phenomena in condensed matter
literature.\re{LQ2}
(Interestingly enough, the free energy density used to describe the
isotropic-nematic transition, the Landau-de Gennes free energy, is formally
identical to the effective potential obtained for the
electroweak transition.)
Here,
we would like to investigate this possibility within the context of hot field
theories.
The point is that the usual vacuum decay
mechanism for first order transitions relies on results obtained
within homogeneous nucleation calculations, which assume that below $T_c$
the system has
small fluctuations about the metastable phase. In other words,
it is assumed that the path integral controlling the transition rate is
dominated by its saddle point, given by the solution to the Euclidean
equations of motion. Small fluctuations are then incorporated by  evaluating
the path integral by a Gaussian approximation. But as the transition
grows weaker,
large amplitude fluctuations about equilibrium will become more probable and
the approximations used will break down.\re{GR} Instead of expanding about a
homogeneous metastable background, one should be expanding about an
inhomogeneous background consisting of these large amplitude sub-critical
fluctuations. This is clearly a very hard task which nevertheless must be
undertook if we are to better
understand the dynamics of weak first order transitions.
The present work is but a first step in this direction
as we attempt to obtain the equilibrium distribution
of sub-critical bubbles at temperatures above $T_c$. This way we can obtain
the fraction of the total volume which is occupied by the
broken symmetric phase as $T_c$ is approached from above, and thus examine
the validity of the homogeneous nucleation picture as a function of the
strength of the phase transition.
Gleiser, Kolb and Watkins \re{GKW} (GKW) have first suggested
this possibility,
concentrating on correlation volume bubbles, but they did not take into account
that subcritical bubbles are unstable and, therefore, shrink.  (See, however,
their comments in Section 3.3). Later, Anderson \re{CRITICS} remarked  that
subcritical bubbles may disappear not only by shrinking but also
due to thermal
noise. As a result of any of these two mechanisms he concluded that
the fraction of the total
volume filled by bubbles of the unstable phase will be much smaller than
proposed by GKW. Here we incorporate these two disappearance
mechanisms into a kinetic description of the transition in order to
investigate the importance of
sub-critical bubbles near the critical temperature. We find that
sub-critical fluctuations become progressively more
important as the strength of the transition weakens, until the point where
the approximations we use in order to treat the problem
analytically break down
and our analysis cannot be trusted quantitatively. However, we believe
that our results are indicative of the relevance of pre-transitional
phenomena in the study of sufficiently weak
phase transitions in hot field theories.

In the next Section we will obtain the equation governing the kinetics of
sub-critical bubbles for temperatures just below $T_1$. In Section 3 we obtain
the expressions for the thermal nucleation rates needed to solve the kinetic
equation. In Section 4 we solve the kinetic equation in three
regimes; assuming
that  the shrinking  dominates the disappearance of subcritical bubbles,
assuming that thermal noise dominates, and finally neglecting
both shrinking and thermal noise.
Starting with the whole volume occupied only by the stable
phase, we obtain the equilibration time scale and the equilibrium number
density of bubbles of the unstable phase in the three  regimes at the
particular temperature considered, and establish the conditions for each
of the three processes to dominate the kinetics. In Section 5 we compare the
three time scales in the context of the standard electroweak model.
As expected,
for strong enough transitions shrinking or thermal noise dominate and the
sub-critical bubbles play a negligible role during the phase transition.
As the transition weakens we find that a larger fraction of the total
volume is occupied by the broken-symmetric phase. The
approximations we used to analytically solve the kinetic
equation break down when a large fraction of the volume is
occupied by sub-critical bubbles of the broken-symmetric phase,
and we
cannot carry our study into the limit of very weak transitions.
But our
results clearly suggest that for weak enough transitions a departure of the
usual vacuum decay mechanism is to be expected. Concluding remarks are
presented in Section 6.

\vspace{24pt}
\thesection{\bf 2. KINETIC EQUATION}
\setcounter{section}{2}
\setcounter{equation}{0}
\vspace{18pt}

Let $\n$ be the number density of sub-critical bubbles of radius $R$ at time
$t$. Thus, ${{\partial\n}\over {\partial R}} dR$
is the number per unit volume of bubbles of radii between $R$ and
$R+dR$. Since the bubbles can shrink, their radius $R$ is a function of time,
$R(t)$. We will only consider bubbles with $R\geq \xi$, where $\xi$ is the
correlation length for fluctuations around equilibrium, given by
\beq
\label{eq:XI}
\xi^{-2}(T)=V^{\prime \prime}(\f=0,T).
\eeq
Bubbles with $R\sim \xi$ will be statistically dominant since any larger
fluctuation has larger free-energy and is exponentially suppressed. (Recall
that for $T\geq T_c$ the free energy of bubbles is a monotonically increasing
function of $R$.\re{GKW})

Consider a large volume $V$ filled by the stable phase $\rf=0$, which is cooled
down from high temperatures to a temperature just below $T_1$. Bubbles with
radius $R$ of the new phase with $\rf=\f_+$ will be thermally nucleated in the
background of the phase $\rf=0$ with number density $\n$; the bubbles are
energetically unfavored and will shrink away with some velocity $dR/dt$.
Shrinking will always be present, unless there is a stabilizing mechanism for
the bubbles, as in non-topological solitons. \re{NTS} [We would like to stress
though that not much is known about nonlinear bubble collapse.
Naively, one would expect small, unstable bubbles to collapse in a time
$\sim R$. In fact,
recent results on the evolution of unstable bubbles found a remarkable
{\it enhancement} of the lifetime of bubbles by 3 to 4 orders of magnitude,
as long as their initial radius
is larger than about $2.5\xi$ and their amplitude at the center probes the
nonlinearity of the potential.\re{PULSONS}]
Thermal noise
may also destroy small sub-critical bubbles.
The importance of this effect will depend both on the ratio of bubble size to
thermal length ($R/T^{-1}$), and
on the
strength of the coupling of the bubbles with the thermal background, which
we parameterize with a dimensionless coefficient $a$.
Anderson\re{CRITICS} wrote this ``thermal destruction''
rate as $ \simeq a T (RT)^3$.  This expression
would imply that
the rate of disappearance of a bubble due to thermal noise
increases with its size, a result we find counter-intuitive.
Instead, we will conservatively take this rate to be
of order $a T$, and thus independent of bubble size. As we are mostly
interested in small bubbles here,
we expect that this overestimate of the true rate will not
compromise our results. In any case, it will only make our final results
stronger, as the final fraction of volume occupied by sub-critical bubbles
will be larger than what we obtain.
This expression yields a lower rate than that of Ref. \ref{CRITICS}, since
 $T > R^{-1}$.
We take, therefore, the rate per unit volume
of disappearance of sub-critical bubbles due to thermal noise to be
$\Gamma_{TN} (R) \simeq a
T/\frac{4}{3} \pi R^3$.

We now proceed to obtain the rate equation.
We can say, quite generally, that the number of bubbles that at time $t+dt$
have radius $R+dR$ in a volume $V\gg\xi$ is equal to the number of bubbles at
time $t$ with radius $R$ plus the net change in the number of bubbles in the
range ($R$, $R+dR$)  due to thermal nucleation,
disappearance due to thermal noise
and shrinking in the time interval $dt$,
\begin{eqnarray}
n(R+dR,t+dt)V= \n V+\left ({{V_0}\over V}\right )\Gamma_{0\rightarrow +}(R)Vdt
 \nonumber\\  -\left ({{V_+}\over V}\right )\Gamma_{+\rightarrow 0}(R)Vdt
- \left(\frac{V_+}{V}\right) \Gamma_{TN} (R) Vdt.
\end{eqnarray}

Thus, to first order in $dR$ and $dt$ we obtain, after
subtracting $n(R+dR,t)$ to both terms and dividing by $Vdt$,
\begin{eqnarray}
\label{eq:KIN}
{{\partial \n}\over {\partial t}}=-{{\partial \n}\over {\partial R}}
\left ({{dR}\over {dt}}\right )+\left ({{V_0}\over V}\right )\Gamma_{0
\rightarrow +}(R)  \nonumber\\
 - \left ({{V_+}\over V}\right )\Gamma_{+\rightarrow 0}(R)
- \left(\frac{V_+}{V}\right) \Gamma_{TN} (R)
\end{eqnarray}
Here, $\Gamma_{0\rightarrow +}(R)$ ($\Gamma_{+\rightarrow 0}(R)$)
is the rate per unit volume for
the thermal nucleation of a bubble of radius $R$ of phase $\f=\f_+$ within
the phase $\f=0$ (phase $\f=0$ within
the phase $\f_+$).
These were the only two rates considered in GKW.\re{GKW}
The volume ratios $V_{0(+)}/V$ take into account the
fact that the total volume in each phase changes in time due to the evolution
of $\n$. The initial conditions we choose are
\beq
V_0(t=0)=V~~~~~~~~{\rm and}~~~~~~~~V_+(t=0)=0~~,
\eeq
that is, all volume $V$ is initially in the phase $\f=0$. Also, $V_+$
must be understood as the volume of the (+)--phase in bubbles of radius $R$
{\it only}, since we are following the evolution of $\n$.
 Thus $\left(
\frac{V_+}{V}\right) = \frac{4}{3} \pi R^3 n (R,t)$.
In Eq.~(\ref{eq:KIN})  bubbles of radius $R$ can disappear due to their
shrinking (accounted for by the first term in the right-hand side with
$dR/dt<0$), due to thermal noise (last term in the right-hand side),
and due
to the nucleation of bubbles of the (0)--phase in their interior.
This latter process will, in general, involve bubbles of (0)--phase of
different radii as we will discuss later on.

Because
\beq
\left ({{V_0}\over V}\right )=1 - \int_{\xi}^{\infty}dR\left[-{4\over 3}\pi
R^3{{\partial\n}\over {\partial R}}\right]~~,
\eeq
Eq.~(\ref{eq:KIN})  is an integro-differential equation and must be solved
within certain approximations, as with any Boltzmann-like equation. In a
complete treatment of the kinetics we should also write an equation for the
fluctuating regions of the (0)--phase within the (+)--phase. Furthermore, there
would be two contributions to $V_0$, a connected background volume
and a bubble-like, disconnected volume. The topology of the two-phase system
will, in general, be very complicated. Thus, we choose the range of
temperatures in which it is consistent to restrict the (0)--phase to the
background, as will be clear later.

Before we move on, we comment on two other possible contributions to $\n$ which
we will not discuss here; a)Induced nucleation: small bubbles may act as seeds
for the nucleation of other bubbles in their neighborhood due to the gain in
surface energy. There should be an enhancement of the nucleation
rate due to the presence of small bubbles in the background,
very much like the presence of impurities in condensed matter systems.
b)Collision of
bubbles: small bubbles may acquire a thermal velocity and collide forming
larger bubbles. Of course, for very short lived bubbles none of these processes
should be very important. However, we again stress that large enough bubbles
can be quite long-lived, and that a truly realistic scenario of weak first
order transitions will be much more complicated than our simple model.
The worth of the present effort relies on it being the first attempt to go
beyond GKW, by
incorporating more complicated out-of-equilibrium processes in the usual
description of phase transitions in field theories.

\vspace{24pt}
\thesection{\bf 3. THERMAL NUCLEATION RATES}
\setcounter{section}{3}
\setcounter{equation}{0}
\vspace{18pt}

In order to solve Eq.~(\ref{eq:KIN}) we need to determine the thermal
nucleation rates $\Gamma_{0(+)\rightarrow +(0)}$. For temperatures $T\la T_1$,
it is reasonable to assume that most bubbles will be nucleated with radius
$R\ga \xi$, where $\xi$ was defined
in Eq.~(\ref{eq:XI}).
For simplicity we will assume the same correlation
length for fluctuations around both minima, even though at $T_1$ the
correlations about $\f_+$ diverge. Following GKW we write for the rates,
\beq
\label{eq:RATES}
\Gamma_{0(+)\rightarrow +(0)}(R)=AT^4{\rm exp}\left [-F\left ({\bar \f}_{+(0)}
\right )/T\right ]~~~,
\eeq
where $A$ is a constant of order unity and the  ansatz for the sub-critical
bubble configurations ${\bar \f}$ is
\beq
\label{eq:BUB}
{\bar \f}_+(r)=\f_+{\rm exp} \left [-r^2/R^2\right ]~~~;~~~
{\bar \f}_0(r)=\f_+\left (1-{\rm exp}\left [-r^2/R^2\right ]\right ),
\eeq
with the free energy functional $F({\bar \f})$ of the bubble configuration
given by
\beq
\label{eq:FREE}
F\left ({\bar \f}\right )=4\pi\int r^2dr\left [{1\over 2}\left ({{d{\bar \f}}
\over {dr}}\right )^2 + V\left ({\bar \f},T\right )\right ]~~.
\eeq
In order to move on, we will choose a specific (but quite general) potential,
\beq
\label{eq:POT}
V(\f,T)={{m^2(T)}\over 2}\f^2 - \gamma(T)\f^3 +{{\l(T)}\over 4}\f^4~~,
\eeq
where $\gamma(T)$ and $\l(T)$ are positive definite functions of $T$ and
$m^2(T)$ can be negative below a certain temperature $T_2<T_c$. This potential
has a minimum at $\f=0$ as long as $m^2(T)>0$, and a local minimum at
$\f_+=\left[3\gamma(T)+\sqrt{9\gamma^2(T)-4m^2(T)\l(T)}~\right]/2\l(T)$, which
appears at a temperature $T_1$ given by the solution of $\gamma^2(T_1)=
4m^2(T_1)\l(T_1)/9$, when $\f_+(T_1)=3\gamma(T_1)/2\l(T_1)$. Below $T_c$
this minimum becomes the global minimum and the minimum at
$\f=0$ becomes metastable.
Using Eqs.
(\ref{eq:BUB}) and (\ref{eq:POT}) in Eq. (\ref{eq:FREE}), we obtain for the
free energy $F({\bar \f})$,\re{GK1}
\beq
\label{eq:FREE1}
F\left ({\bar \f}_{0(+)}\right)=\alpha_{0(+)}R + \beta_{0(+)}R^3~~,
\eeq
where
\beq
\label{eq:ALPHA}
\alpha_0=\alpha_+={{3\sqrt{2}}\over 8}\pi^{3/2}\f_+^2~~,
\eeq

\ba
\label{eq:BETA0}
\beta_0 & = &\pi^{3/2}\f_+^2\left[ {{m^2(T)}\over 8}(\sqrt2-8)
+\gamma(T)\f_+\left(3-{{3\sqrt2}\over 4}+{{\sqrt3}\over 9}\right)\right.
   \nonumber\\
& & \left.
+\l(T)\f_+^2\left(-1+{{3\sqrt2}\over 8}-{{\sqrt3}\over 9}+{1\over {32}}
\right)\right]~~,
\ea
and
\beq
\label{eq:BETA+}
\beta_+=\pi^{3/2}\f_+^2\left [{{m^2(T)\sqrt2}\over 8}-
{{\sqrt{3}}\over 9}\gamma(T)\f_+~
+{{\l(T)}\over {32}}\f_+^2\right ]~~.
\eeq

\vspace{24pt}
\thesection{\bf 4. SOLVING THE KINETIC EQUATION}
\setcounter{section}{4}
\setcounter{equation}{0}
\vspace{18pt}

We will solve Eq.~(\ref{eq:KIN}) in different regimes assuming that
the dominant process for the disappearance of bubbles is: 1.) their
shrinking, 2.) thermal noise and 3.) the  nucleation of bubbles of the
 true vacuum inside them.
This will allow us to make a direct comparison
of the results for the relaxation time scale and equilibrium number
density obtained in these regimes, in the context of the electroweak
transition in the next Section.
\vspace{16pt}

\centerline{\bf 4.1 Kinetic Equation with Shrinking}

We include only the shrinking term, neglecting the last two terms
of Eq.~(\ref{eq:KIN}), and solve this equation with the approximation,
\beq
\label{eq:CONS}
\left({{V_0}\over V}\right)=1 - \int_{\xi}^{\infty}dR\left[-{4\over 3}\pi
R^3{{\partial\n}\over {\partial R}}\right]\equiv 1-|I|\simeq 1~~.
\eeq
In words, we will solve the kinetic equation in the regime in which
 the volume in the (+)--phase, given by the sum
over all bubbles with
radii from $\xi$ to $\infty$ (actually, to $R\sim V^{1/3}\gg \xi$), is
small, i.e. $|I|\ll 1$. We
will check our results for consistency after obtaining $n(R,t)$.
With this approximation the
kinetic equation becomes,
\beq
\label{eq:KINS}
{{\partial\n}\over {\partial t}} = {{\partial\n}\over {\partial R}}
\left[f(R)\right] + \Gamma_{0\rightarrow +}(R)~~,
\eeq
where we wrote $dR/dt=-f(R)<0$. The positive definite function $f(R)$
encompasses the physically interesting cases $dR/dt=v={\rm constant}$, and
$dR/dt\sim 1/R$. We enforced the shrinking explicitly by choosing the minus
sign in $dR/dt$. The rate $\Gamma_{0\rightarrow +}$ was defined in
Eq.~(\ref{eq:RATES}), with the free energy given by Eq.~(\ref{eq:FREE1}) with
$\alpha_+$ and $\beta_+$ given by Eqs.~(\ref{eq:ALPHA}) and (\ref{eq:BETA+})
respectively. In order to solve this equation analytically we make one further
approximation; we neglect the volume term in the expression for the free
energy, so that the nucleation rate can be written as
\beq
\Gamma_{0\rightarrow +}(R)\simeq AT^4{\rm exp}\left[-\alpha_+R/T\right]\equiv
g(R)~~.
\eeq
Here we have defined the function $g(R)$. Given that most bubbles have radii
$R\ga \xi$ this should be a good approximation
that
must be tested in each application. For the electroweak case, independently
of the parameters of the model, we obtain
$\alpha_+ \xi/\beta_+ \xi^3 \simeq 6.65$ for $T=T_1$. The ratio increases for
lower temperatures.

In order to solve the above equation, first note that the time
independent equilibrium
number density at the temperature chosen, ${\bar n}(R)$, is the solution of
\beq
\label{eq:EQ}
{{\partial {\bar n}(R)}\over {\partial R}}= -{{g(R)}\over {f(R)}}~~,
\eeq
which is easily obtained as
\beq
{\bar n}(R)={\bar n}(\xi) - \int_{\xi}^R{{g(R)}\over {f(R)}}dR~~.
\eeq
Choosing $f(R)=v$ and the definition of $g(R)$ above we obtain,
\beq
\label{eq:nEQ1}
{\bar n}(R)={{AT^5}\over {v\alpha_+}}{\rm exp}\left[-\alpha_+R/T\right] ~~.
\eeq
Using Eq.~(\ref{eq:EQ}), we can rewrite the kinetic equation as
\beq
\label{eq:Y}
{{\partial Y(R,t)}\over {\partial t}}={{\partial Y(R,t)}\over {\partial R}}
f(R),
\eeq
where we introduced the departure from equilibrium, $Y(R,t)\equiv \n - {\bar n}
(R)$. This equation is solved by writing
\beq
Y(R,t)=X(R){\rm exp}\left[-t/\tau_1\right]~~,
\eeq
so that $\tau_1$ is the relaxation time and the solution approaches the
equilibrium distribution as $t\rightarrow \infty$. Choosing again $f(R)=v$ we
obtain upon substitution in Eq.~(\ref{eq:Y}),
\beq
\label{eq:X}
X(R)=X_0{\rm exp}\left[-R/v\tau_1\right]~~,
\eeq
where $X_0$ is a constant, which, together with $\tau_1$, will
be fixed by the initial condition $Y(R,0)=-
{\bar n}(R)$. Using Eq.~(\ref{eq:nEQ1}) we find
\beq
\label{eq:TAU1}
X_0=-{{AT^5}\over {v\alpha_+}}~~~{\rm and}~~~\tau_1={T\over {v\alpha_+}}~~.
\eeq
Putting the results together we find for the solution of the kinetic equation,
\beq
\label{eq:SOL1}
\n={\bar n}(R)\left(1-e^{-t/\tau_1}\right)={{AT^5}\over {v\alpha_+}}
e^{-\alpha_+R/T}\left[1 - e^{-\alpha_+vt/T}\right]~~.
\eeq
Thus, the relaxation time to approach the equilibrium distribution
${\bar n}(R)$ is the
time for a bubble of characteristic radius $v\tau_1=T/\alpha_+$ to shrink with
constant velocity $v$. (The reason why $\tau_1$ is independent of $R$ is due to
our choice of constant shrinking speed for all bubbles.) Substituting this
solution into the consistency condition given in Eq.~ (\ref{eq:CONS}) we obtain
\beq
\label{eq:CONS1}
|I|={4\over 3}\pi \xi^3{\bar n}(\xi)\left[1+3\left({T\over {\alpha_+\xi}}
\right) + 6\left({T\over {\alpha_+\xi}}\right)^2 + 6\left({T\over {\alpha_+
\xi}}\right)^3\right]\ll 1~.
\eeq
The consistency condition is then an expression of the validity of the
semi-classical approximation, which requires that exp $ \left [- F\left ({\bar
\f}_+ \right )/T\right ]\simeq $
exp $ \left [- \alpha_+R/T \right ] \ll 1$; if the
smallest bubble has a large nucleation barrier ($=F_+/T$) its production is
exponentially suppressed and only a negligible fraction of the total volume
will be occupied by the (+)--phase.

\vspace{16pt}
\centerline{\bf 4.2 Kinetic Equation with Thermal Destruction}

In Ref.~\ref{CRITICS} Anderson writes the thermal destruction rate
per unit volume as
\begin{equation}
\Gamma{(R)}_{TN} \simeq T^4 ~.
\end{equation}
With this expression, the rate of disappearance of a bubble increases with its
 volume. As mentioned above, we think this is counter-intuitive, since larger
 bubbles should be less prone to be detroyed by thermal noise than smaller
ones,
  not the contrary. In lack of a better understood expression we will take this
 rate to be $a T$, on dimensional grounds, independently of the size of the
 bubble.
The rate per unit volume of the disappearance of bubbles
due to thermal noise is then,
\begin{equation}
\Gamma{(R)}_{TN} = \frac{aT}{{{4 \pi}\over 3} R^3}~~.
\end{equation}
 The constant $a$ is proportional to the coupling of
the bubble to the thermal bath to some power.
 This is a difficult quantity to obtain
microscopically, and will depend on how the field couples to itself and
to other fields in the model. There has been some progress recently
in the computation of the viscosity coefficient in hot field theories,
although the results are model dependent.\re{MORIK} However, it is by now
clear that
viscosity (which is related to the coupling to the bath by
the fluctuation-dissipation theorem) in field theories is at least
a two-loop effect.

The kinetic equation is now,
\begin{equation}
\frac{\partial n(R,t)}{\partial t}  = \left(\frac{V_0}{V}\right)
\Gamma_{0 \to +} (R) - \frac{4}{3} \pi R^3 n (R, t)~ \Gamma_{TN} (R)~~.
\end{equation}
We will again make the approximation of Eq. (4.1). Using  Eqs. (4.3)
 and (4.13), the kinetic equation becomes
\begin{equation}
\frac{\partial n(R,t)}{\partial t}  = g (R) - a T n(R,t)~~.
\end{equation}
Thus, the time independent equilibrium number density is now
\begin{equation}
\bar{n} (R) = \frac{g(R)}{aT}={{AT^3}\over {a}}
e^{-\alpha_+R/T}~~,
\end{equation}
and we solve the equation for ($n - \bar{n}$) with the ansatz
\begin{equation}
n (R, t) - \bar{n} (R) = X(R) e^{- t/\tau_2}~~,
\end{equation}
and the initial condition $n (R,0) = 0$, which yields $X (R) = \bar{n}
(R)$.  Thus, $\tau_2^{-1} = aT$ and
\begin{equation}
n(R,t) = \bar{n} (R) ( 1 -  e^{-a Tt})={{AT^3}\over {a}} e^{-\alpha_+R/T}
( 1 - e^{-a Tt})
\end{equation}
The relaxation time is thus $\tau_2 = (aT)^{-1}$ as
expected.  Comparing this value with the relaxation time when shrinking
alone is considered $\tau_1 \simeq T/\alpha_+ v$, as obtained in the
previous section, we see that, under the approximations we choosed,
thermal noise will be dominant, {\it i.e.}
$\tau_2 < \tau_1$, for
\begin{equation}
a > \frac{\alpha_+ v}{T^2} \simeq \frac{\phi_+^2 v}{T^2} .
\end{equation}
Here we have taken  $\alpha_+ \simeq \phi_+^2$ from Eq. (3.6).
This is an important condition, since, as we mentioned
above, $a$ is proportional to some
power  of the coupling of the field $\phi$ with the thermal bath,
and $\phi_+/T$ is a measure of the strength of the
transition. (For successful baryogenesis during the electroweak scale
one needs roughly $\phi_+/T\ga 1$.\re{REVBAR})
Thus, thermal noise only dominates the dynamics over shrinking for
sufficiently weak transitions, {\it and} for strong enough
coupling to the thermal
bath.
The consistency  condition  in Eq. (4.1) is  again given by
Eq.  (4.12), but in this case ${\bar n}(\xi)$ is given  by Eq. (4.17). Thus,
this condition  is again an expression of the  validity of  the semiclassical
approximation, as explained below Eq. (4.17).

\vspace{16pt}
\centerline{\bf 4.3 Kinetic Equation Without Shrinking and Thermal
Destruction}

Neglecting the shrinking of bubbles and their coupling to thermal noise,
the kinetic equation is
\beq
{{\partial \n}\over {\partial t}}=\left({{V_0}\over V}\right)\Gamma_{0
\rightarrow +}(R) - \left({{V_+}\over V}\right)\Gamma_{+\rightarrow 0}~~,
\eeq
where the nucleation rates have been defined before. In order to be
consistent with our previous approach, we will again take $V_0/V\simeq 1$.
Since $\left(V_+/V \right) \Gamma_{+\rightarrow 0}(R)$
is the rate at which bubbles of (+)--phase of radius $R$ disappear due to the
nucleation of a region of (0)--phase in its interior,
as mentioned before, $V_+$ is the total
volume of all bubbles of radius $R$ (and not the total volume of (+)--phase).
 Thus we
have $V_+/V=(4\pi/3)R^3\n $. Consistency  with the assumption $V_0/V\simeq 1$
as $t\rightarrow \infty$
requires that  $(4\pi/3)R^3{\bar n}(R)\ll 1$. Since
most bubbles of the (+)--phase will have $R\simeq \xi$, this condition is
essentially the same as that of Eq.~(\ref{eq:CONS1}), as it should.

In principle, to obtain $\Gamma_{+\rightarrow 0}$ we should sum over bubbles of
the (0)--phase  of different radii which can trigger the disappearance of
regions of the (+)--phase.  This sum should be dominated by the contribution of
bubbles of radius $\xi$. In this case the dominant process for wiping out
regions of (+)--phase is nucleation of correlation volume regions of the
(0)-phase, as was explicitly assumed in the work of GKW. However, whenever we
will need to write $\Gamma_{+\rightarrow 0}$ explicitly to obtain analytic
expressions, we will approximate it with the rate of
thermal nucleation of bubbles of radius $R$,  $\Gamma_{+\rightarrow 0} \simeq
\Gamma_{+\rightarrow 0} (R)$. We expect this approximation to be good since
most bubbles of (+)--phase have radii $R$ not much larger than $\xi$ anyway.
The kinetic equation is, therefore,
\beq
\label{eq:KIN2}
{{\partial \n}\over {\partial t}}=\Gamma_{0\rightarrow +}(R) -
{{4\pi}\over 3}R^3\n \Gamma_{+\rightarrow 0}~~.
\eeq
We can rewrite it as,
\beq
\label{eq:Z}
{{\partial z(R,t)}\over {\partial t}}= - q(R)z(R,t),~~~~
z(R,t)\equiv p(R)-q(R)\n ~,
\eeq
where we defined
\beq
p(R)\equiv \Gamma_{0\rightarrow +}(R)~~~{\rm and}~~~
q(R)\equiv {{4\pi}\over 3}R^3\Gamma_{+\rightarrow 0}~~~.
\eeq
The solution to Eq.~(\ref{eq:Z}) is simply
$z(R,t)=z_0{\rm exp}[-q(R)t]$, or, using the initial condition
$n(R,0)=0$,
\beq
\label{eq:SOL2}
\n = {\bar n}(R)\left[1 - e^{-q(R)t}\right]~~;~~{\bar n}(R)={{p(R)}\over
{q(R)}}~~.
\eeq
For a given radius $R$, the relaxation time is then
\beq
\label{eq:TAU2}
\tau_3(R) = \left[q(R)\right]^{-1} =
\left[{{4\pi}\over 3}R^3\Gamma_{+\rightarrow 0}\right]^{-1}~~,
\eeq
that is, the time scale for the nucleation of a bubble of the (0)--phase inside
a bubble of radius $R$ of the (+)--phase. Note that $\tau_3$ depends
exponentially on $R^3$; bubbles of larger radii will approach the equilibrium
distribution at a much slower rate than smaller bubbles. The fastest bubbles to
equilibrate are those with $R \simeq \xi$.
With the approximation $\Gamma_{+\rightarrow 0} \simeq
\Gamma_{+\rightarrow 0} (R)$ and
using the expressions of Section 2 for the nucleation rates,
the equilibrium distribution is given by
\beq
\label{eq:N2}
{{4\pi}\over 3}R^3{\bar n}(R)={\rm exp}\left[-{{\left(\beta_+-\beta_0\right)
R^3}\over T}\right]~~.
\eeq
The consistency condition then states that the volume occupied by the
(+)--phase is small,
\beq
\label{eq:CONS2}
{{4\pi}\over 3}\xi^3{\bar n}(\xi)={\rm exp}\left[-{{\left(\beta_+-\beta_0
\right)\xi^3}\over T}\right] \ll 1~~.
\eeq

\vspace{24pt}
\thesection{\bf 5. APPLICATION TO THE ELECTROWEAK TRANSITION}
\setcounter{section}{5}
\setcounter{equation}{0}
\vspace{18pt}

For a given model with a first order transition which of the approaches
above better describes the dynamics of thermal fluctuations? We are
limited to study the transition at $T\la T_1$ when only small sub-critical
fluctuations exist. However, even for this limited temperature range we
should be able to examine the importance of incorporating the shrinking of
the bubbles and their coupling to thermal noise into the description of the
kinetics. In this Section we will do this in the context of the standard
electroweak model using the 1-loop approximation to the effective potential.
Even though recent work has shown that the 1-loop approximation is not adequate
due to large infrared corrections at $T_c$,\re{IR} we will take the 1-loop
potential as an example of how to apply our methods. Application to other
potentials is quite straightforward.

We will compare the different relaxation time scales obtained above,
$\tau_1$, $\tau_2$ and
$\tau_3$, and investigate which mechanism for the suppression of regions of
broken-symmetric phase will dominate as the strength of the transition
changes. Our strategy is as follows. First we compare $\tau_1$ with $\tau_2$,
that is, we compare the relaxation time scale incorporating only shrinking
(Section 4.1) with the relaxation time scale incorporating only thermal
noise (Section 4.2). We have already shown under our assumptions
that thermal noise dominates
over shrinking if the parameter $a$ satisfies the inequality
$a > \alpha_+v/T^2$, Eq. (4.20). Since $a$ is a free parameter in our model
we have two possibilities, depending on the value of $a$. If $a$ does not
satisfy the inequality, shrinking dominates and thermal noise is a sub-dominant
mechanism during the approach to equilibrium. Otherwise, shrinking is
the sub-dominant mechanism. Since we then will know
the values of $a$ for which shrinking or thermal noise dominates as a function
of the strength of the transition, we can
proceed by comparing these two processes
with the inverse
thermal nucleation rate time scale, given by $\tau_3$ (Section 4.3). As a
result, we will be able to establish which mechanism dominates the dynamics as
the transition's strength varies for different values of $a$.

We take the Higgs mass to be the parameter that controls  the strength of the
transition,  while we fix the top mass at $m_T=130$ GeV.
For the 1-loop approximation to the electroweak potential, we can write\re{AH}
\beq
\label{eq:VEW}
V_{\rm EW}(\f,T)=D\left(T^2-T_2^2\right)\f^2-ET\f^3+{{\l_T}\over 4}\f^4~~,
\eeq
where the constants $D$ and $E$ are given by $D=\left[6(m_W/\s)^2+
3(m_Z/\s)^2+6(m_T/\s)^2\right]/24$, and $E=\left[6(m_W/\s)^3
+3(m_Z/\s)^3\right]/12\pi\simeq
10^{-2}$. $T_2$ is the temperature at which the origin becomes unstable,
given by
\beq
T_2=\sqrt{\left(m_H^2-8B\s^2\right)/4D}~~,
\eeq
where the physical Higgs mass is given in terms of the 1-loop corrected
$\l$ as $m_H^2=\left(2\l+12B\right)\s^2$, with $B=\left(6m_W^4+3m_Z^4-12m_T^4
\right)/64\pi^2\s^4$. We use $m_W=80.6$ GeV, $m_Z=91.2$ GeV, and $\s=246$ GeV.
The temperature corrected Higgs self-coupling is
\beq
\l_T=\l- {1\over {16\pi^2}}\left[\sum_Bg_B\left({{m_B}\over {\s}}\right)^4
{\rm ln}\left(m_B^2/c_BT^2\right)-\sum_Fg_F\left({{m_F}\over {\s}}\right)^4
{\rm ln}\left(m_F^2/c_FT^2\right)\right],
\eeq
where the sum is performed over bosons and fermions (in our case only the
top quark) with their respective degrees of freedom $g_{B(F)}$.
Also, ${\rm ln}~c_B=5.41$ and ${\rm ln}~c_F=2.64$.

This potential is equivalent to the potential of Eq.~(\ref
{eq:POT}), with the replacements $m^2(T)=2D(T^2-T_2^2)$, $\gamma(T)=ET$, and
$\l(T)=\l_T$. The temperatures $T_1$ and $T_c$ are given by
\beq
\label{eq:T}
T_1=T_2/\sqrt{1-9E^2/8\l_TD},~~{\rm and}~~ T_c=T_2/\sqrt{1-E^2/\l_TD}~~.
\eeq

${\rm From}$ these values for $T_1$ and $T_c$, it is clear that the
quantity $x\equiv E^2/\l_TD$ can be used as a measure of the strength of
the transition. For $x=0$, the transition is second order. As shown in
Ref.~\ref{GK2}, $x\simeq 0.027$ at $m_H=60$ GeV, becoming smaller for larger
Higgs masses. The transition is already quite weak at the experimental lower
bound of $m_H=57$ GeV.\re{H} In Fig. 1 we show
the quantity $x$ as a function of
the Higgs mass for $m_H\ge 35$ GeV.

With the potential above, we can easily compute the nucleation rates using the
expressions of Section 2, as done by Gleiser and Kolb.\re{GK1} We have then all
that is needed to compute the relaxation time scales $\tau_1$, $\tau_2$, and
$\tau_3(\xi)$ [since $\tau_3$ depends on $R$, we consider only the fastest
bubbles to equilibrate, those of radius $\sim \xi$],
and the equilibrium distributions ${\bar n}(R)$.

Let us start by comparing the relaxation time scales for the shrinking term
and the thermal destruction term, following Eq. (4.20). In Fig. 2 we plot
$\alpha_+(T_1)/T^2_1$ (we took $v=1$)
as a function of the Higgs mass. Thermal noise
dominates over shrinking
if $a$ lies above the curve, for a given Higgs mass. Notice that
for small Higgs masses, {\it i.e.} strong transitions, shrinking should be
the dominant process unless the coupling to the bath is unrealistically
large (or non-perturbative). However, since there are uncertainties in the
precise expression for the thermal destruction rate, we will consider both
possibilities, with shrinking and thermal noise dominating the dynamics,
respectively.
If shrinking dominates, we can neglect thermal noise and compare $\tau_1$
and $\tau_3(\xi)$ directly. This is done in the continuous lines of
Fig. 3 where
we show the ratio of $\tau_1/\tau_3(\xi)$ as a function of the
Higgs mass. The results are shown for two different shrinking velocities,
$v=1$ and $v=0.4$.
For both velocities, with Higgs masses below $52$ GeV we find that
$\tau_1 \ll \tau_3(\xi)$, so that the approach to equilibrium
is dominated by the shrinking.
In Fig. 4 we show the two consistency conditions
Eqs.~(\ref{eq:CONS1}) and (\ref{eq:CONS2}) as a
function of the Higgs mass. We see that  the fraction of the total
volume occupied by the (+)--phase increases rapidly as the
Higgs mass increases. That is, as the transition grows weaker a larger
fraction of the volume is occupied by the broken-symmetric phase.
Even though our results indicate
that we cannot trust our approximations for $m_H\ga 53$
GeV, where $V_+/V\ga 10\%$, it is quite clear that the reason our
approximations break down is precisely the large
fraction of the volume occupied
by sub-critical bubbles. These results suggest that pre-transitional
phenomena will be present for weak enough transitions and that we should
expect modifications of the vacuum decay picture in this case. A
precise description of the transition is beyond the scope of the present
formalism.

We now consider the case in which thermal noise dominates over shrinking.
By inspecting Fig. 2, we can find an appropriate value of $a$ which is large
enough for the inequality Eq. (4.20)
to be always satisfied. We will take $a=0.5$ as an
illustration, even though this value is probably unrealistically large.
In Fig. 3 the dashed line shows the ratio $\tau_2/\tau_3(\xi)$
as a function of the Higgs mass. Note that again thermal noise becomes
sub-dominant as the Higgs mass is increased, although it does so at a
slower rate than the shrinking term. However, the qualitative conclusion is
the same in both cases. For large enough Higgs masses, the fraction of the
total volume in the broken-symmetric phase becomes substantial due to the
increasing weakness of the transition. This weakness is characterized by
the fact that both the shrinking of sub-critical bubbles as well as their
destruction due to thermal noise become less important in the description
of the transition. Since the breakdown of our approximations is due to the
failure of the semi-classical, or dilute gas approach, we are confident that
as the Higgs mass increases beyond the limit of validity of our approximations
a regime will be reached in which a departure from the usual false vacuum decay
mechanism is to be expected, although at this point we cannot afford
to make a quantitative prediction. A rough estimate may be obtained by
comparing, at the nucleation temperature,
the typical distance between sub-critical bubbles to the radius
of a critical bubble obtained by the usual calculations.
If the distance between sub-critical bubbles is of the order of
the critical radius, the usual nucleation mechanism must be revised.
Work on this topic is currently in progress.

\vspace{24pt}
\thesection{\bf 6. CONCLUDING REMARKS}
\vspace{18pt}

We have obtained a kinetic equation describing the approach to equilibrium in
first-order transitions for temperatures above the critical temperature. For
sufficiently strong transitions
and making approximations that we consider reasonable,
 we were able to analytically solve the kinetic
equation for three
different regimes, determined by the dominant mechanism responsible for
the suppression of regions of the broken-symmetric phase within the
symmetric phase. The three processes are the shrinking of the sub-critical
bubbles, their destruction due to the thermal bath to which they couple, and
by thermal nucleation of regions of the symmetric phase in their interior.
By obtaining the relaxation time-scales in all regimes, we were able to study
the relative importance of each process in the early stages of the
transition.

We applied our approach to the standard electroweak transition, showing that
for Higgs masses below 55 GeV or so, the approach to
equilibrium is dominated by
either shrinking or thermal destruction, depending on the strength of the
coupling of bubbles to the thermal bath. In this case, the total volume
occupied by the equilibrium distribution of sub-critical bubbles of the
broken-symmetric phase is negligible, and the transition proceeds by the
usual vacuum decay mechanism.
As the Higgs mass increases and the transition
becomes progressively weaker, we found that a larger fraction of the total
volume becomes occupied by the broken-symmetric phase, forcing the breakdown
of our analytical approximations. However, it is clear from our results that
a regime will eventually be reached in which a substantial fraction of the
volume is in the broken-symmetric phase as the critical temperature is
reached. In this case we should expect a departure from the usual vacuum
decay mechanism. We cannot be quantitative about the
value of the Higgs
mass (for a given top mass)
where this occurs due to the limitations of our analytical approach.
Given that the lower bound on the Higgs mass is now above 60 GeV, we expect
very interesting physics to be lurking behind our present knowledge of the
dynamics of weak first-order transitions.

\vspace{36pt}

\centerline{ \bf ACKNOWLEDGEMENTS}
\vspace{18pt}

We thank E. W. Kolb and D. Seckel for useful discussions. We would like to
thank the Institute for Theoretical Physics at Santa Barbara for its
hospitality during the Cosmological Phase Transitions program, where this
work was initiated.
(GG) was supported in part by the Department of Energy under the contract
No. DE-FG03-91ER 40662, Task C.
(MG) was supported in part by National Science Foundation grants
No. PHY-9204726 at Dartmouth and No. PHY-8904035 at
the Institute for Theoretical
Physics at Santa Barbara.

\vspace{1.0in}
\centerline{{\bf References}}
\frenchspacing
\begin{enumerate}

\item\label{NUC} For reviews see, J. D. Gunton, M. San Miguel, and P. S. Sahni,
in {\it Phase Transitions and Critical Phenomena}, edited by C. Domb and J. L.
Lebowitz (Academic, London, 1983), Vol. 8; J. S. Langer, Lectures presented
at l'\'Ecole de Physique de la Mati\`ere Condens\'ee, Beg Rohu, France,
August 1989, ITP preprint NSF-ITP-90-221i.

\item\label{LAN} See the review by J. Langer in Ref. 1.

\item\label{A} E.W. Kolb and M.S. Turner, {\it The Early Universe},
(Addison-Wesley, 1990); A.D. Linde, {\it Particle Physics and Inflationary
Cosmology}, (Harwood Academic Publishers, 1990).

\item\label{LQ} T. Stinton III and J. Lister, {\it Phys. Rev. Lett.}
{\bf 25}, 503 (1970). H. Zink and W.H. de Jeu, {\it Mol. Cryst. Liq. Cryst.}
{\bf 124}, 287 (1985). For other condensed matter systems exhibiting
pre-transitional phenomena see, for example, L.A. F\'ernandez, J.J.
Ruiz-Lorenzo, M. Lombardo, and A. Taranc\'on, {\it Phys. Lett. B} {\bf  277},
485 (1992); A. Gonzalez-Arroyo, M. Okawa, and Y. Shimizu, {\it Phys. Rev.
Lett.} {\bf 60}, 487 (1988).

\item\label{LQ2} J.J. Stankus, R. Torre, C.D. Marshall, S.R. Greenfield,
A. Sengupta, A. Tokmakoff, and M.D. Fayer, {\it Chem. Phys. Lett.} {\bf 194},
213 (1992).

\item\label{GR} M. Gleiser and R.O. Ramos, {\it Phys. Lett. B} {\bf 300},
271 (1993).

\item\label{GKW} M. Gleiser, E. W. Kolb, and R. Watkins, {\it Nucl. Phys.}
{\bf B364}, 411 (1991); M. Gleiser, {\it Phys. Rev.} {\bf D42}, 3350 (1990).

\item\label{CRITICS} G. Anderson, {\it Phys. Lett. B} {\bf 295}, 32 (1992);
See also,
M. Dine, R. Leigh, P. Huet, A. Linde, and D. Linde,
{\it Phys. Rev. D} {\bf 46}, 550 (1992).

\item\label{NTS} J. Frieman, G. Gelmini, M. Gleiser, and E. W. Kolb,
{\it Phys. Rev. Lett.} {\bf 60}, 2101 (1988).

\item\label{PULSONS} M. Gleiser, Dartmouth College report No. DART-HEP-93/05,
August 1993; V.G. Makhankov, {\it Phys. Rep. C} {\bf 35}, 1 (1978).

\item\label{GK1} M. Gleiser and E. W. Kolb, {\it Phys. Rev. Lett.}
{\bf 69}, 1304 (1992).

\item\label{MORIK} A. Hosoya and M. Sakagami, {\it Phys. Rev. D} {\bf 29},
2228 (1984); A. Hosoya, M. Sakagami,and M. Takao, {\it Ann. Phys.} (NY)
{\bf 154}, 229 (1984); M. Morikawa, {\it Phys. Rev. D} {\bf 33}, 3607 (1986).

\item\label{REVBAR} A.G. Cohen, D.B. Kaplan, and A.E. Nelson, UC San Diego
report No. UCSD-PTH-93-02, to appear in
{\it Ann. Rev. Nucl. Part. Sci.}, {\bf 43}.

\item\label{IR} G. Boyd, D. Brahm and S.D.H. Hsu, Harvard preprint No.
HUTP-92-A027 (1992), in press {\it Phys. Rev. D};
M. E. Carrington, {\it Phys. Rev. D} {\bf 45},
2933 (1992); M. Dine {\it et. al.} in Ref.~\ref{CRITICS};
P. Arnold, {\it Phys. Rev.} {\bf D46}, 2628 (1992); M. Gleiser
and E. W. Kolb, {\it Phys. Rev. D} {\bf 48}, 1560 (1993).

\item\label{AH} G.W. Anderson and L.J. Hall, {\it Phys. Rev.} {\bf D45},
2685 (1992).

\item\label{H} ALEPH, DELPHI, L3 and OPAL Collaborations, as presented by
M. Davier, Proceedings of the International Lepton-Photon Symposium and
Europhysics Conference on High Energy Physics, eds. S. Hegerty, K. Potter and
E. Quercigh,  November 1991, to appear.

\item\label{GK2} M. Gleiser and E. W. Kolb, in Ref.~\ref{IR}.

\end{enumerate}

\newpage

FIG.~1. The parameter $x=E^2/\l_TD$ as a function of the Higgs mass for
several values of the top mass.

\vspace{1.0in}

FIG.~2 The ration of relaxation time-scales for shrinking and thermal
destruction processes as a function of the Higgs mass. The region above
the curve defines the domain for thermal destruction domination.

\vspace{1.0in}

FIG.~3. The continuous lines denote the
ratio between the shrinking and reverse nucleation time-scales,
as a function
of the Higgs mass for shrinking velocities $v=1$ and $v=0.4$.
The dashed line denotes the ratio between thermal destruction and reverse
nucleation time-scales with parameter $a=0.5$.
\vspace{1.0in}

FIG.~4. The consistency conditions of Eqs.~\ref{eq:CONS1} and \ref{eq:CONS2}
as a function of the Higgs mass.

\end{document}